\documentstyle[twocolumn,pra,aps,floats,epsf,lscape,grafik]{revtex}
\begin{document}
\draft

\twocolumn[\hsize\textwidth\columnwidth\hsize\csname@twocolumnfalse\endcsname
\vspace{-1cm}
\begin{flushright}
Phys. Rev. A, in press
\end{flushright}

\title
{\bf Bose-Einstein condensation in a two-dimensional, trapped, 
interacting gas}
\author{M. Bayindir and B. Tanatar}
\address{
Department of Physics,
Bilkent University,
Bilkent, 06533 Ankara, Turkey}
\maketitle
\begin{abstract}
We study Bose-Einstein condensation phenomenon in a
two-dimensional (2D) system of bosons subjected to an harmonic
oscillator type confining potential. The interaction among the
2D bosons is described by a delta-function in configuration
space. Solving the Gross-Pitaevskii equation within
the two-fluid model we calculate the condensate fraction, ground
state energy, and specific heat of the system. Our results
indicate that interacting bosons have similar behavior to those
of an ideal system for weak interactions.
\end{abstract}

\vskip1pc]

\narrowtext
\newpage

The observation of the Bose-Einstein condensation (BEC) phenomenon
in dilute atomic gases\cite{1,2,3,4} has caused a lot of
attention, because it provides opportunities to study the 
thermodynamics of weakly interacting systems in a controlled
way. The condensate clouds obtained in the experiments consist of 
a finite number of atoms (ranging from several thousands to
several millions), and are confined in an externally applied
confining potentials. The ground state properties of the
condensed gases, including the finite size effects on the
temperature dependence of the condensate fraction, are of
primary interest. At zero temperature, the mean-field
approximation provided by the Gross-Pitaevskii equation\cite{5}
describes the condensate  rather well and at finite temperatures
a self-consistent Hartree-Fock-Bogoliubov (HFB) approximation is
developed.\cite{6} Path Integral Monte Carlo (PIMC)
simulations\cite{7} on three-dimensional, interacting bosons 
appropriate to the current experimental conditions demonstrate 
the effectiveness of the mean-field type approaches. Various aspects 
of the mean-field theory, as well as detailed calculations
corresponding to the available experimental conditions are 
discussed by Giorgini {\it et al}.\cite{8}

In this work, we examine the possibility of BEC in a
two-dimensional (2D) interacting atomic gas, under a trap potential.
Such a system may be realized by making one dimension of the
trap very narrow so that the oscillator states are largely
separated. Possible experimental configurations in spin
polarized hydrogen and magnetic waveguides are currently under
discussion.\cite{9}
The study of 2D systems is also interesting theoretically,
since even though the homogeneous system of 2D bosons do not 
undergo BEC,\cite{10} number of examples\cite{11} have indicated such a
possibility upon the inclusion of confining potentials.
We employ the two-fluid, mean-field model developed by Minguzzi
{\it et al}.\cite{12} to study the 2D Bose gas. Similar 
approaches\cite{13} are gaining attention because of their simple and 
intuitive content which also provide semi-analytical expressions 
for the density distribution of the condensate. Recently,
Mullin\cite{14} considered the self-consistent mean-field theory
of 2D Bose particles interacting via a contact interaction
within the Popov and semi-classical approximations. His
conclusions were that a phase transition occurs for a 2D Bose
system, in the thermodynamic limit, at some critical temperature, 
but not necessarily to a Bose-Einstein condensed state. However,
in the current experiments the finite number of atoms $N$
prevent various divergences to give rise to behavior akin to
non-interacting systems.  

Our work is motivated by the success of mean field, two-fluid
models\cite{12,13} {\it vis {\` a} vis} more involved calculations
and direct comparison with experiments.
In the following we briefly describe the two-fluid model of Minguzzi
{\it et al}.\cite{12} and present our results for the 2D Bose
gas.

The condensate wave function $\Psi(r)$ is described by the
Gross-Pitaevskii (GP) equation\cite{5}

\begin{eqnarray}
-{\hbar^2\over 2m}\,\nabla^2\Psi(r)+V_{\rm ext}(r)\Psi(r)+2gn_1(r)
&&\Psi(r)+g\Psi^3(r) \nonumber \\ 
&& =\mu\Psi(r)\, ,
\end{eqnarray}

where $g$ is the repulsive, short-range interaction strength, 
$V_{\rm ext}(r)= m\omega^2r^2/2$ is the confining (or trap) potential, 
$n_1(r)$ is 
the distribution function for the non-condensed particles. We note that 
unlike in a three-dimensional system, $g$ in our case is not
simply related to the $s-$wave scattering length, but will be
treated as a parameter. In the two-fluid model developed by
Minguzzi {\it et al}.\cite{12} the non-condensed particles are
treated as bosons in an effective potential
$V_{\rm eff}(r)=V_{\rm ext}(r)+2gn_1(r)+2g\Psi^2(r)$, and having 
the same chemical potential $\mu$ with that of the condensate. The
density distribution is given by
\begin{equation}
n_1(r)=\int {d^2p\over (2\pi\hbar)^2}\,{1\over
\exp{\{(p^2/2m+V_{\rm eff}(r)-\mu)/k_BT\}}-1}\, ,
\end{equation}
and the chemical potential is fixed by the relation
\begin{equation}
N=N_0+\int {\rho(E)dE\over \exp{[(E-\mu)/k_BT]}-1}\, ,
\end{equation}
where $N_0=\int \Psi^2(r)\,d^2r$, is the number of condensed
atoms, and the semi-classical density of states is calculated 
using\cite{12,15,16}
\begin{equation}
\rho(E)={m\over 2\pi\hbar^2}\,\int_{V_{\rm eff}(r)<E} d^2r\, .
\end{equation}

\bfig{t}\ff{0.4}{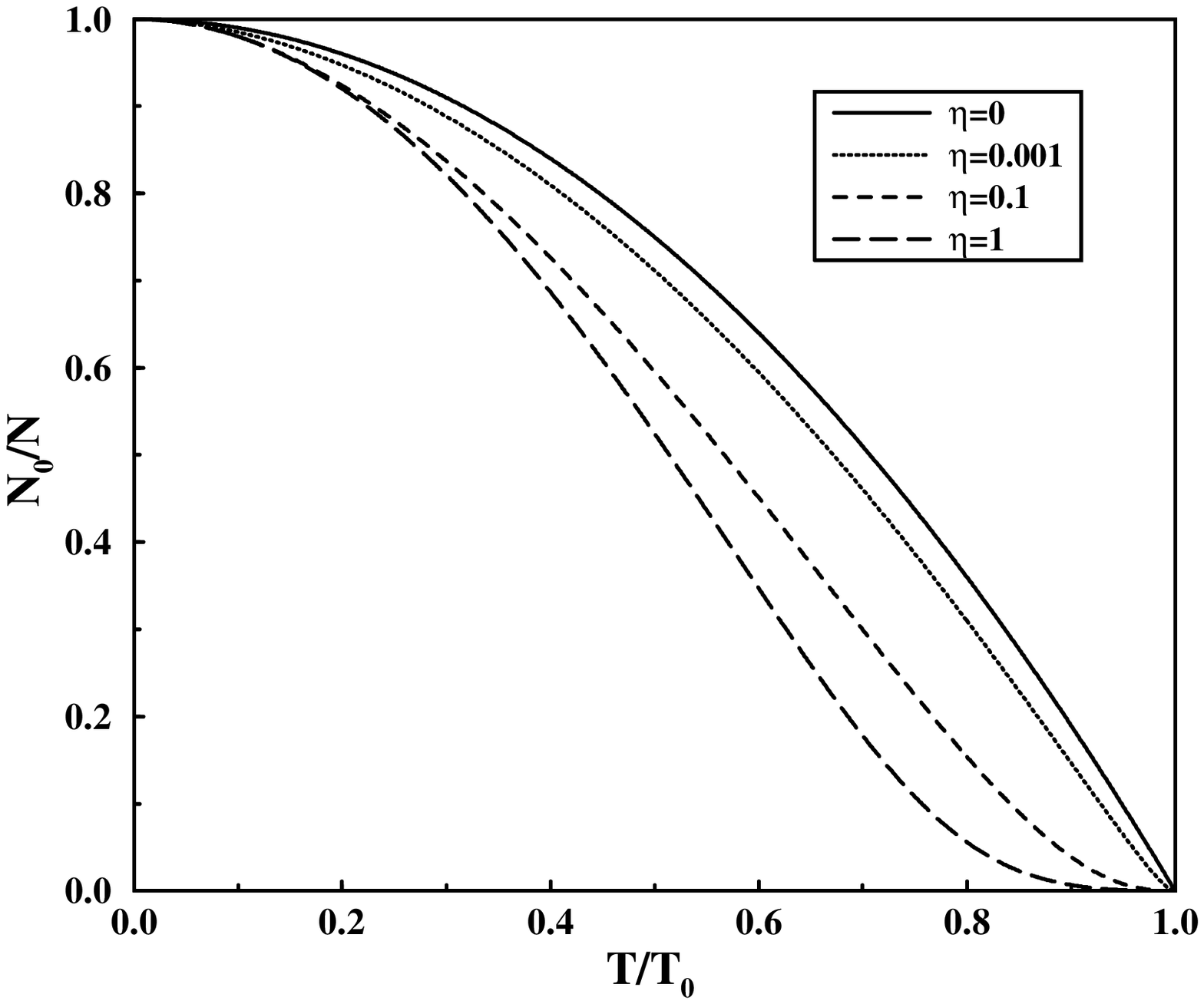}\efig{The condensate fraction $N_0/N$ as a
 function of $T/T_0$, for a system of $N=10^5$ atoms. The various interaction 
strengths are described by the parameter $\eta$}{}

The GP equation admits a simple solution within the Thomas-Fermi
approximation, i.e. when the kinetic energy term is neglected,
\begin{equation}
\Psi^2(r)={1\over g}[\mu-V_{\rm ext}(r)-2gn_1(r)]\theta(\mu-V_{\rm
ext}(r)-2gn_1(r))\, ,
\end{equation}
where $\theta(x)$ is the unit step-function. Thomas-Fermi
approximation is regarded to be rather good except for the
region close to the phase transition.\cite{17} Minguzzi {\it et
al}.\cite{12} have numerically solved the above set of equations
self-consistently. They have also introduced a simpler approximation
scheme which treats the interaction effects perturbatively.
Encouraged by the success of even the zero-order solution in
describing the fully numerical self-consistent solution in the
3D case, we attempt to look at the situation in 2D.
In a similar vein, we treat the interactions among the 
non-condensed particles perturbatively. To zero order in 
$gn_1(r)$, the number of condensed particles are calculated to be
\begin{equation}
N_0={\pi\hbar^2\over g m}\,\left({\mu\over\hbar\omega}\right)^2\, ,
\end{equation}
and the density of states is obtained as
\begin{equation}
\rho_0(E)=\left\{ 
\begin{array}{ll}
E/(\hbar\omega)^2 & \quad\hbox{if}\quad \mu<0\, ,\\
&\\
2(E-\mu)/(\hbar\omega)^2 & \quad\hbox{if}\quad 2\mu>E \quad
(\mu>0)\, ,\\
&\\
E/(\hbar\omega)^2 & \quad\hbox{if}\quad 2\mu<E \quad (\mu>0)\, .
\end{array}
\right.
\end{equation}
If we use the above form of the density of states, valid for $E>0$,
then we obtain
\begin{equation}
N=N_0+t^2\left[{\pi^2\over
3}-\hbox{dilog}{\left(1-e^{-\alpha/t}\right)}\right]\, ,
\end{equation}
where $t=k_BT/\hbar\omega$, and $\alpha=\mu/\hbar\omega$. The
chemical potential $\mu(N,T)$ is obtained as the solution of
this transcendental equation.

\bfig{t}\ff{0.4}{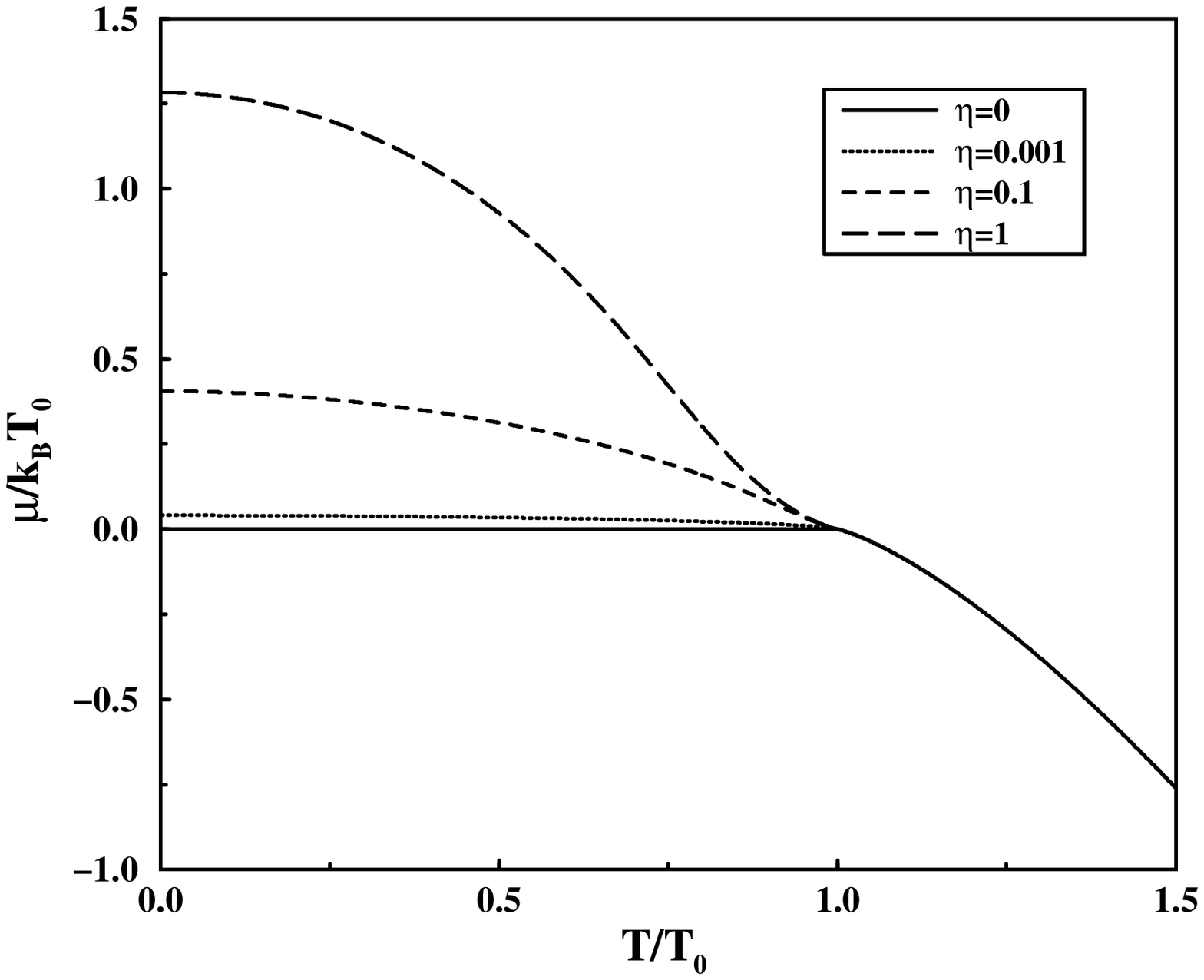}\efig{The temperature dependence of the chemical potential $\mu$ for various interaction strengths.}{}

In Fig.\,1 we show the temperature dependence of the condensate
fraction $N_0/N$ for a system of $N=10^5$ particles, and for
various values of the interaction strength.
Also shown for comparison is the result for an ideal 2D Bose
gas in an harmonic trap, given by $N_0/N=1-(T/T_0)^2$ where
$k_BT_0=\hbar\omega(N/\zeta(2))^{1/2}$. 
We observe that BEC-like behavior occurs for small values of the 
parameter
$\eta=mg/\pi\hbar^2$, i.e. the weakly interacting system. Here
we identify the BEC with the macroscopic occupation of the
ground state at $T=0$ and the depletion of it above $T_0$. As the 
strength of interactions is increased we find that the temperature
dependence of $N_0/N$ deviates from the non-interacting case more
noticeably. Mullin\cite{14} has argued that there is no BEC in
2D in the thermodynamic limit. We consider a system with finite 
number of particles, and 
we were able to obtain a self-consistent solution for the
chemical potential for various values of the interaction
strength as displayed in Fig.\,2.
We next evaluate the temperature dependence of the internal
energy $\langle E\rangle=[{\langle E\rangle}_{\rm
nc}(N-N_0)/2+{\langle E\rangle}_{\rm c}]/N$ which consists of 
contributions from the non-condensed particles

\begin{eqnarray}
{\langle E\rangle}_{\rm nc} & = &k_BT_0\,\left({\zeta(2)\over
N}\right)^{1/2}\,\left[\right.{\pi^2\alpha t^2\over
3}+t^3 (2\zeta(3) \nonumber\\ 
&& +\int_0^{\alpha/t} {x^2dx\over
e^x-1})-\alpha^2 t\ln{\left(1-e^{-\alpha/t}\right)}\left. \right]\,
,
\end{eqnarray}

and the condensed particles
\begin{equation}
{\langle E\rangle}_{\rm c}=k_BT_0\,{1\over
3\eta}\,\left({\zeta(2)\over N}\right)^{1/2}\,\alpha^3\, .
\end{equation}
In the above expressions $\zeta(n)$ is the Riemann zeta-function.
The kinetic energy of the condensed particles is neglected in
accordance with our Thomas-Fermi approximation to the GP
equation.
In Fig.\,3, we display the temperature dependence of $\langle E\rangle$
for different values of the interaction strength. The
non-interacting energy is simply
$\langle E\rangle/Nk_BT_0=[\zeta(3)/\zeta(2)](T/T_0)^5$. 
For small $\eta$,
and $T<T_0$, the behavior of $\langle E\rangle$ resembles to that in 
a 3D system. As $\eta$ increases, a bump in $\langle E\rangle$ 
develops for $T<T_0$, which perhaps indicates the breakdown of the 
present approximation or an artifact of the calculation. We have
no physical explanation for this behavior.
The corresponding results for the specific heat
$C_V=d\langle E\rangle/dT$, are shown in Fig.\,4. In contrast to the
non-interacting case where a sharp peak at $T=T_0$ is seen, the
effects of short-range interactions smoothes out the transition.
However, this smoothing is partly due to the finite number of
particles in the system.\cite{8} The effects of interactions and finite
number of particles are not disentangled in our treatment. 

\bfig{t}\ff{0.4}{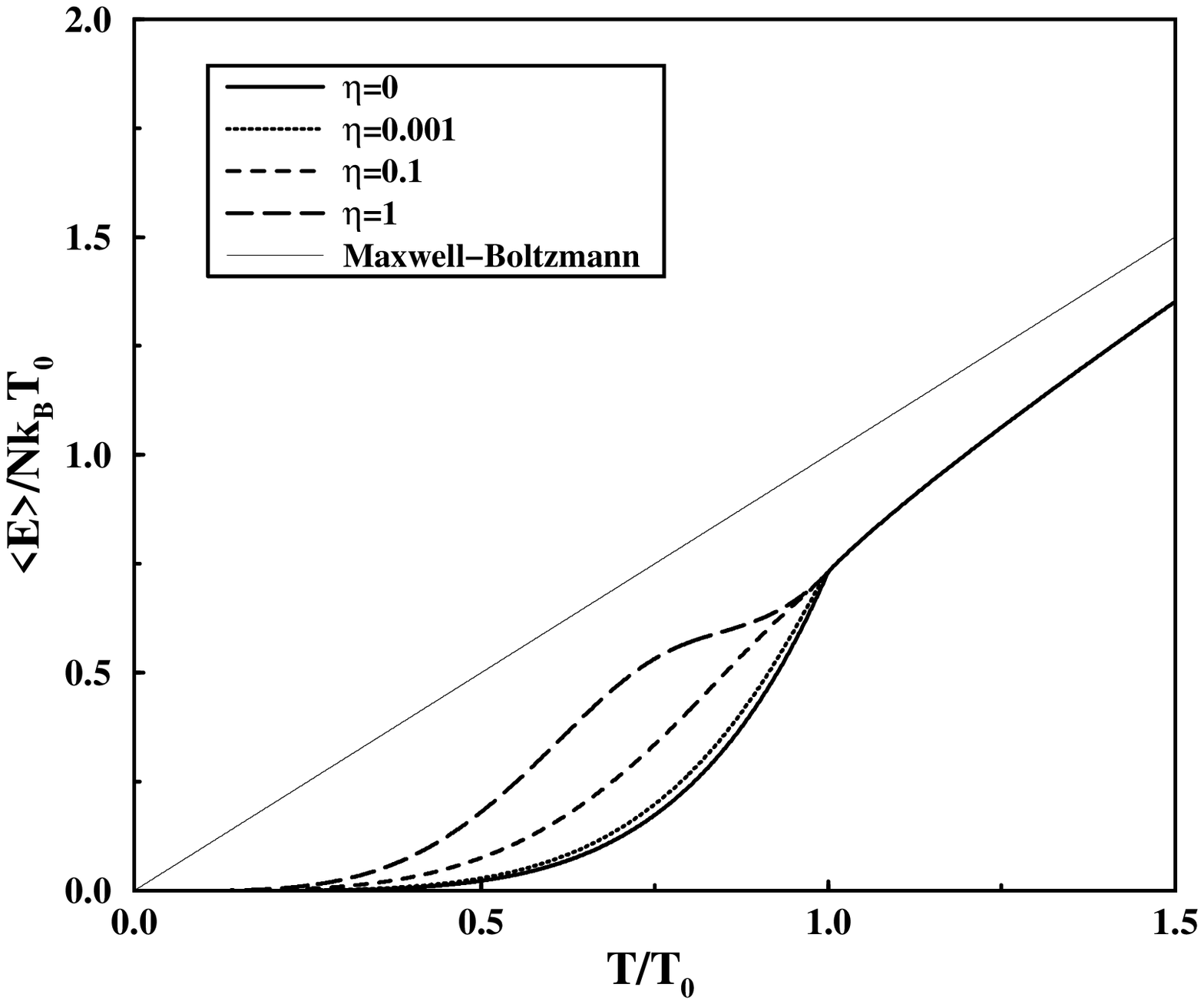}\efig{The ground state energy $\langle E\rangle$ of the 2Dbosons as a function of temperature for various interaction strengths. The Maxwell-Boltzmann result is shown by the thin solid line.}{}

\bfig{t}\ff{0.4}{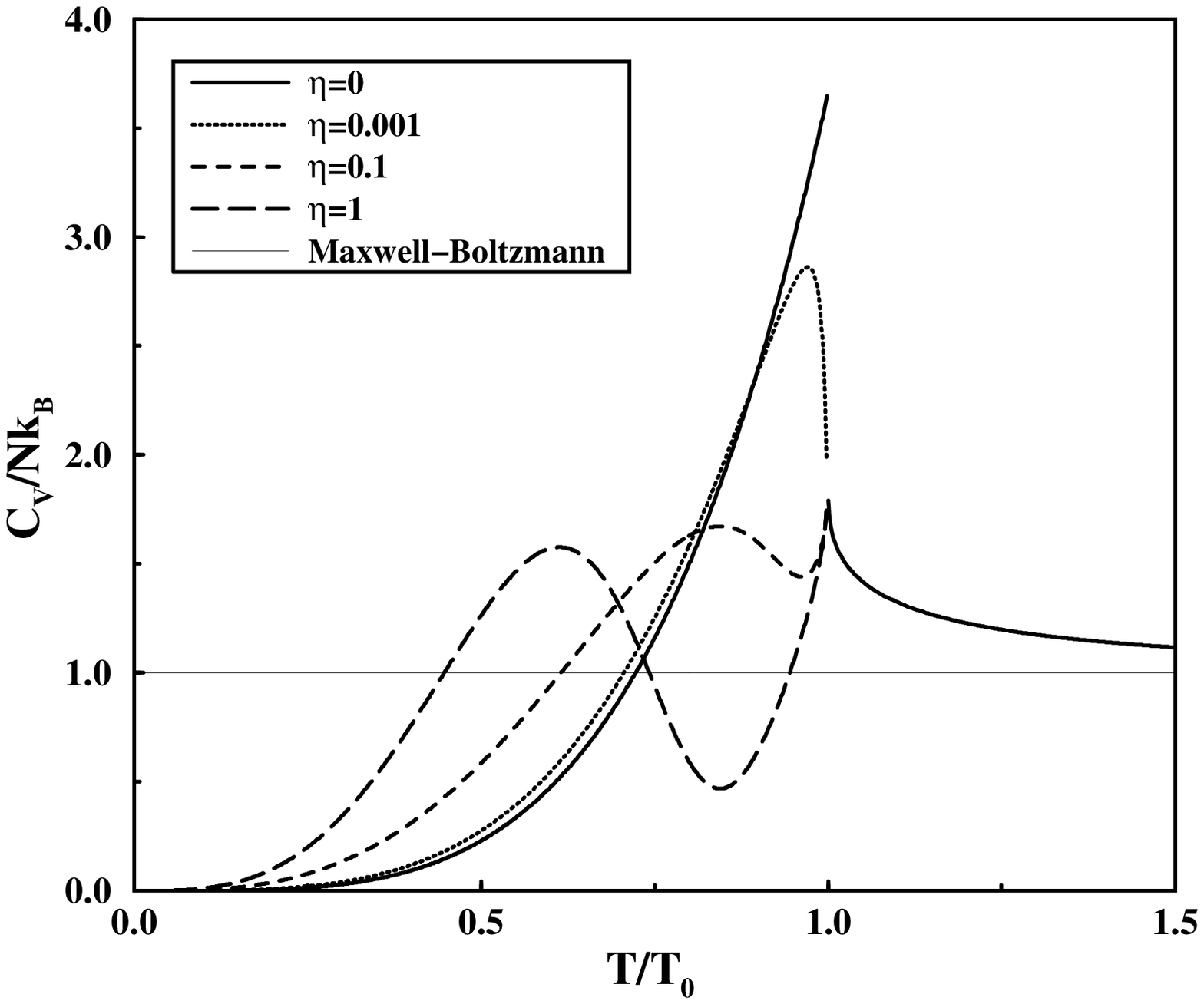}\efig{The specific heat $C_V=d\langle E\rangle/dT$ as a function of temperature for various interaction strengths.}{}

It is a straightforward generalization to include the effects of
anisotropy within the present formalism. For an external
potential of the type $V_{\rm ext}(r)=m\omega^2_x(x^2+\lambda^2
y^2)/2$, where $\lambda=\omega_y/\omega_x$ is the anisotropy
parameter, both $N_0$ and $\rho_0(E)$ depend inversely on $\lambda$.
Similarly, our analysis may be extended to study other power
law potentials such as $V\sim r^\gamma$ for which $\gamma\approx 1$
appears to be interesting.\cite{18,19}

Our calculations using the two-fluid model of Minguzzi {\it et
al}.\cite{12} show that the BEC, in the sense of macroscopic
occupation of the ground state, may occur in a 2D trapped Bose gas when
the short-range interparticle interactions are not too strong.
As the interaction strength increases we could not find self-consistent
solutions to the mean-field equations signaling the breakdown of
our approach. We note that instead of using the lowest order
perturbation approach adopted here, the full solution to the 
self-consistent equations may alleviate the situation. Given the
unclear nature of the phase transition\cite{14,18} in 2D and the 
interest of future experiments, we think it is worthwhile to
perform first-principles calculations.
Recent PIMC simulations\cite{18} on 2D hard-core bosons confirm the
possibility of BEC in the sense that a sharp drop in $N_0/N$
around $k_BT_c\approx 0.78\,N^{1/2}$ is observed for finite
systems.

In summary, we have applied the mean-field, semi-classical
two-fluid model for trapped interacting Bose gases to the case
in two-dimensions. We have found that for a range of interaction
strength parameters the behavior of the thermodynamic quantities
resembles that of non-interacting bosons in a harmonic trap.

{\it Acknowledgments}:
This work was partially supported by the
Scientific and Technical Research Council of Turkey (TUBITAK).
We thank Dr. S. Conti for providing us with the details of his
calculations and Dr. S. Pearson for sending us a preprint of
their work prior to publication. We gratefully acknowledge useful 
discussions with E. Ke{\c c}ecio{\u g}lu, H. Mehrez, and T. Senger.

\end{document}